\documentclass[pra,twocolumn,showpacs,amsmath,amssymb,aps,superscriptaddress,floatfix]{revtex4}
\usepackage{graphicx}
\usepackage{dcolumn}
\usepackage{bm}
\usepackage{ulem}
\newcommand{\ket}[1]{\vert #1 \rangle}

\newcommand{\meanvalue}[3]{\langle #1 \vert #2 \vert #3 \rangle}
\newcommand{\ketbra}[2]{\vert #1 \rangle \langle #2 \vert}
\begin{document}
\preprint{APS/123-QED}
\title{State mapping and discontinuous
entanglement transfer in a multipartite open system}
\author{Matteo Bina}
 \affiliation{Dipartimento di Fisica, Universit\`{a} di Milano, I-20133 Milano, Italy}
 \affiliation{CNISM, UdR Milano, I-20133, Milano, Italy}
\author{Federico Casagrande}%
\affiliation{Dipartimento di Fisica, Universit\`{a} di Milano, I-20133 Milano, Italy}
\affiliation{CNISM, UdR Milano, I-20133, Milano, Italy}
\email{federico.casagrande@mi.infn.it}
\author{Marco G. Genoni}
\affiliation{Dipartimento di Fisica, Universit\`{a} di Milano, I-20133 Milano, Italy}
\affiliation{CNISM, UdR Milano, I-20133, Milano, Italy}
\author{Alfredo Lulli}
\affiliation{Dipartimento di Fisica, Universit\`{a} di Milano, I-20133 Milano, Italy}
\affiliation{CNISM, UdR Milano, I-20133, Milano, Italy}
\author{Matteo G. A. Paris}
\affiliation{Dipartimento di Fisica, Universit\`{a} di Milano, I-20133 Milano, Italy}
\affiliation{CNISM, UdR Milano, I-20133, Milano, Italy}
\affiliation{ISI Foundation, I-10133, Torino, Italy}
\date{\today}
\begin{abstract}
We describe the transfer of quantum information and correlations from an
entangled tripartite bosonic system to three separate qubits through
their local environments also in the presence of various dissipative
effects. Optimal state mapping and entanglement transfer are shown in
the framework of optical cavity quantum electrodynamics involving
qubit-like radiation states and two-level atoms via the mediation of
cavity modes.  For an input GHZ state mixed with white noise we show the
occurrence of sudden death and birth of entanglement, that is
discontinuously exchanged among the tripartite subsystems.
\end{abstract}
\maketitle
\section{Introduction} 
As early as in 1935 Einstein, Podolski, and Rosen \cite{EPR}
as well as Schr\"{o}dinger \cite{Sch} drew the attention on the
correlations in quantum composite systems and the problems raised by
their properties. Much later, theoretical \cite{Bell} and
experimental \cite{Aspect} cornerstones elucidated the issue of
nonlocality. Entanglement is currently viewed as the key resource
for quantum information (QI) processing \cite{Nielsen}, where it
allowed a number of achievements such as teleportation
\cite{Bennett}, cryptography \cite{Gisin} and enhanced measurements
\cite{entame}. 
The deep meaning of multipartite entanglement, its
quantification and detection \cite{Guhne1}, the possible
applications, are the object of massive investigation.
As a matter of fact, optical systems
have been a privileged framework for encoding and manipulating
quantum information, since bipartite and multipartite entanglement
may be effectively generated either in the discrete or continuous
variable regime. On the other hand, the development of QI also
requires localized registers, e.g. for the storage of entanglement
in quantum memories. 
\\\indent Cavity quantum electrodynamics (CQED) \cite{Haroche_Raimond}
is a relevant scenario for this kind of investigations and has been 
addressed for the analysis of entanglement transfer of bipartite
\cite{P04,P204,P304,mem,Hald00,Son02,Zou06,Cas07,CLP2008}
and multipartite \cite{Paternostro,Casagrande,Cirac,Ser06} entanglement.
In this
framework we present a complete study on the entanglement dynamics of a nine
parties open system whose implementation could be feasible in the
optical regime of CQED \cite{Nussmann,ions}. In particular we
describe a system where three radiation modes, prepared in
qubit-like entangled states, are coupled by optical fibers to three
separate optical cavities each of them containing a trapped
two-level atom.  This paradigmatic example allows us investigating
multipartite entanglement transfer and swapping in a more realistic
way than in \cite{Paternostro,Casagrande}, shedding light on
fundamental processes related to quantum interfaces and memories in
quantum networks \cite{Cirac,Ser06}.
\\ \indent 
We demonstrate that a complete mapping of pure entangled
states onto the tripartite atomic subsystem occurs when the external
field has fed the cavities. If this field is then switched off, the
quantum correlations can be periodically mapped onto the tripartite
atomic and cavity mode subsystems according to a triple
Jaynes-Cummings (JC) dynamics \cite{JC}. In the case of external
radiation prepared in a mixed Werner state we deal with the recently
observed phenomenon of entanglement sudden death (ESD) (and birth
(ESB)) \cite{YuPRL,Almeida}, that in our case involves the abrupt
vanishing (and raising) of quantum correlations in tripartite
systems. Though this is in general a still open problem,
nevertheless we can show the occurrence of discontinuous exchange of
tripartite entanglement via ESD effects. We also describe the
dissipative effects introduced by the presence of external
environments, such as the decay of cavity modes, atomic excitations,
and fiber modes. Some results are then reported in the case that the
coupling between external and cavity mode
fields is generalized from monomode to multimode.\\
\indent In Sec. II we introduce the model of the physical system. In
Sec. III we derive all main results concerning state mapping,
entanglement transfer, and entanglement sudden death, also adding
the presence of external environments. The case of multimode
external-cavity field coupling is addressed in Sec. IV. Some
conclusive remarks are reported in Sec. V.
\section{Model of the physical system}
\indent We consider an entangled tripartite bosonic system (f),
prepared in general in a mixed state, interacting with three qubits
(a) through their local environments (c) also in the presence of
dissipative effects. In the interaction picture the system
Hamiltonian has the form:
\begin{eqnarray}
\label{H_int} \hat{\mathcal{H}}^I=\hbar \big\{\sum_{J=A,B,C}\big[
g_J(\hat{c}_J\hat{\sigma}^{\dag}_J+\hat{c}^{\dag}_J\hat{\sigma}_J)\big]+\,\nonumber\\
+\sum_{J,K=A,B,C}\big[\nu_{J,K}(t)(\hat{c}_J\hat{f}^{\dag}_K+\hat{c}^{\dag}_J\hat{f}_K)\big]\big\}
\end{eqnarray}
The operators $\hat{c}_J,\hat{c}^{\dag}_J$
($\hat{f}_J,\hat{f}_J^{\dag}$) are the annihilation and creation
operators for the cavity (external radiation) modes, while
$\hat{\sigma}_J,\hat{\sigma}_J^{\dag}$ are the raising and
lowering operators for the atomic qubits in each subsystem
($J=$A,B,C). We consider real coupling constant $g_J$ for the
atom-cavity mode interaction and $\nu_{J,K}(t)$ for the
interaction of each cavity mode with the three modes of driving
external radiation. We take time dependent constants in order to
simulate the interaction switching-off at a suitable time
$\tau_{off}$ for the external field. Now we take into account
three processes of dissipation: the cavity losses at rate
$\kappa_c$ due to interaction with a thermal bath with a mean
photon number $\bar{n}$, the atomic spontaneous emission with a
decay rate $\gamma_a$ for the upper level, and a loss of photons
inside the fibers at a rate $\kappa_f$. All dissipative effects
can be described under the Markovian limit by standard Liouville
superoperators so that the time evolution of the whole system
density operator $\hat{\rho}(t)$ can be described by the following
ME written in the Lindblad form:
\begin{eqnarray}\label{ME} \dot{\hat{\rho}}&=&-\frac{i}{\hbar}\left
[\hat{\mathcal{H}}_e,\hat{\rho}\right ]+\sum_{J=A,B,C}
\big[\hat{C}_{f,J}\hat{\rho}\hat{C}_{f,J}^{\dag}+\hat{C}^{(g)}_{c,J}\hat{\rho}\hat{C}_{c,J}^{(g)\dag}+\,\nonumber\\
&+&\hat{C}^{(l)}_{c,J}\hat{\rho}\hat{C}_{c,J}^{(l)\dag}+\hat{C}_{a,J}\rho\hat{C}_{a,J}^{\dag}\big]
\end{eqnarray}
where the non-Hermitian effective Hamiltonian is
\begin{eqnarray}\label{He}
\hat{\mathcal{H}}_e&=&\hat{\mathcal{H}}^I-\frac{i\hbar}{2}\sum_{J}\big[\hat{C}^{\dag}_{f,J}\hat{C}_{f,J}+\hat{C}^{(g)\dag}_{c,J}\hat{C}^{(g)}_{c,J}+\,\nonumber\\
&+&\hat{C}^{(l)\dag}_{c,J}\hat{C}^{(l)}_{c,J}+\hat{C}^{\dag}_{a,J}\hat{C}_{a,J}\big].
\end{eqnarray}
The jump operators for the atoms are
$\hat{C}_{a,J}=\sqrt{\gamma_a}\hat{\sigma}_J$, for the fiber
losses $\hat{C}_{f,J}=\sqrt{\kappa_f}\hat{f}_J$, and for the
cavity modes
$\hat{C}_{c,J}^{(l)}=\sqrt{\kappa_c(\bar{n}+1)}\hat{c}_J$ (loss of
a photon) and
$\hat{C}_{c,J}^{(g)}=\sqrt{\kappa_c\bar{n}}\hat{c}^{\dag}_J$ (gain
of a photon). The above ME can be solved numerically by the Monte
Carlo Wave Function method \cite{MCWF}. From now on we consider
dimensionless parameters, all scaled to the coupling constant
$g_A$, and times $\tau=g_At$.\\
\indent As a significative example, the implementation of our scheme
may be realized in the optical regime of CQED by choosing a continuous
variable (CV) entangled field for the subsystem (f) and two-level atoms
as qubits (a). Each qubit is trapped in a one-sided optical cavity (c),
where the radiation modes can be coupled to the cavity modes via optical
fibers as in \cite{SMB}. In optical cavities thermal noise is
negligible ($\bar{n}\cong 0$), spontaneous emission can be effectively
suppressed, and single atoms can remain trapped even for several seconds
\cite{Nussmann}.\\
\indent Here we focus on external field prepared in a qubit-like
entangled state $\hat{\rho}_f(0)$ because this is the condition
for high entanglement transfer for CV field \cite{Casagrande}. The
generation of photon number multimode entangled radiation was
recently demonstrated \cite{Papp}. Under qubit-like behavior we
can describe the entanglement of all three-qubit subsystems (a, c,
f) by combining the information from tripartite negativity
\cite{Sabin}, entanglement witnesses \cite{Acin} for the two
inequivalent classes GHZ and W \cite{Dur}, and recently proposed
criteria for separability \cite{Ghune2}. In fact, the tripartite
negativity $E^{(\alpha)} (\tau)$($\alpha$=a, c, f), defined as the
geometric mean of the three bipartite negativities \cite{Vidal},
is an entanglement measure providing only a sufficient condition
for entanglement detection, though its positivity guarantees the
GHZ-distillability
that is an important feature in QI.
\section{State mapping and tripartite entanglement transfer for single mode coupling}
\subsection{Hamiltonian regime for external field in qubit-like pure
states} 
\begin{figure}[h]
\includegraphics[width=0.23\textwidth]{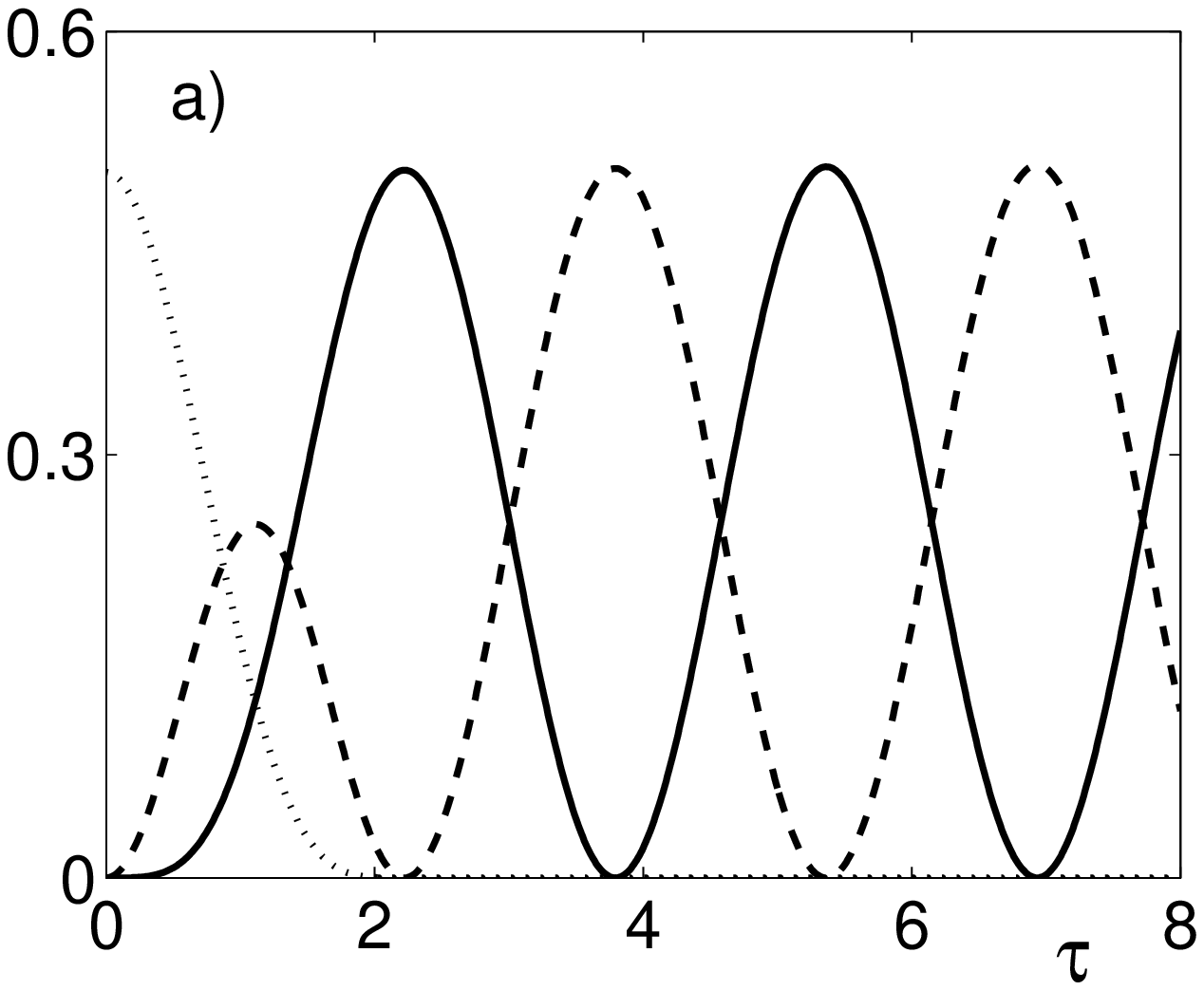}
\includegraphics[width=0.23\textwidth]{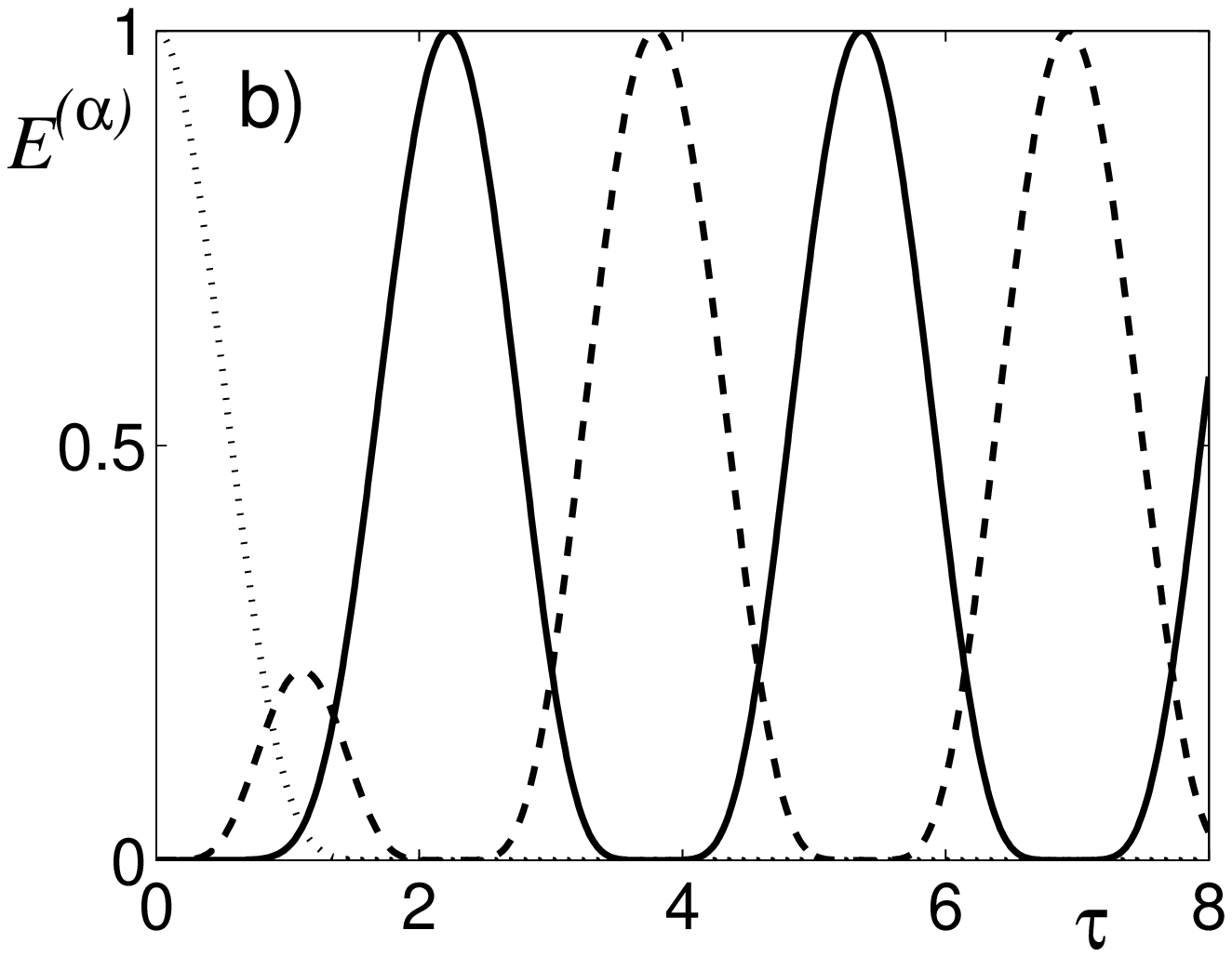}\\
\includegraphics[width=0.23\textwidth]{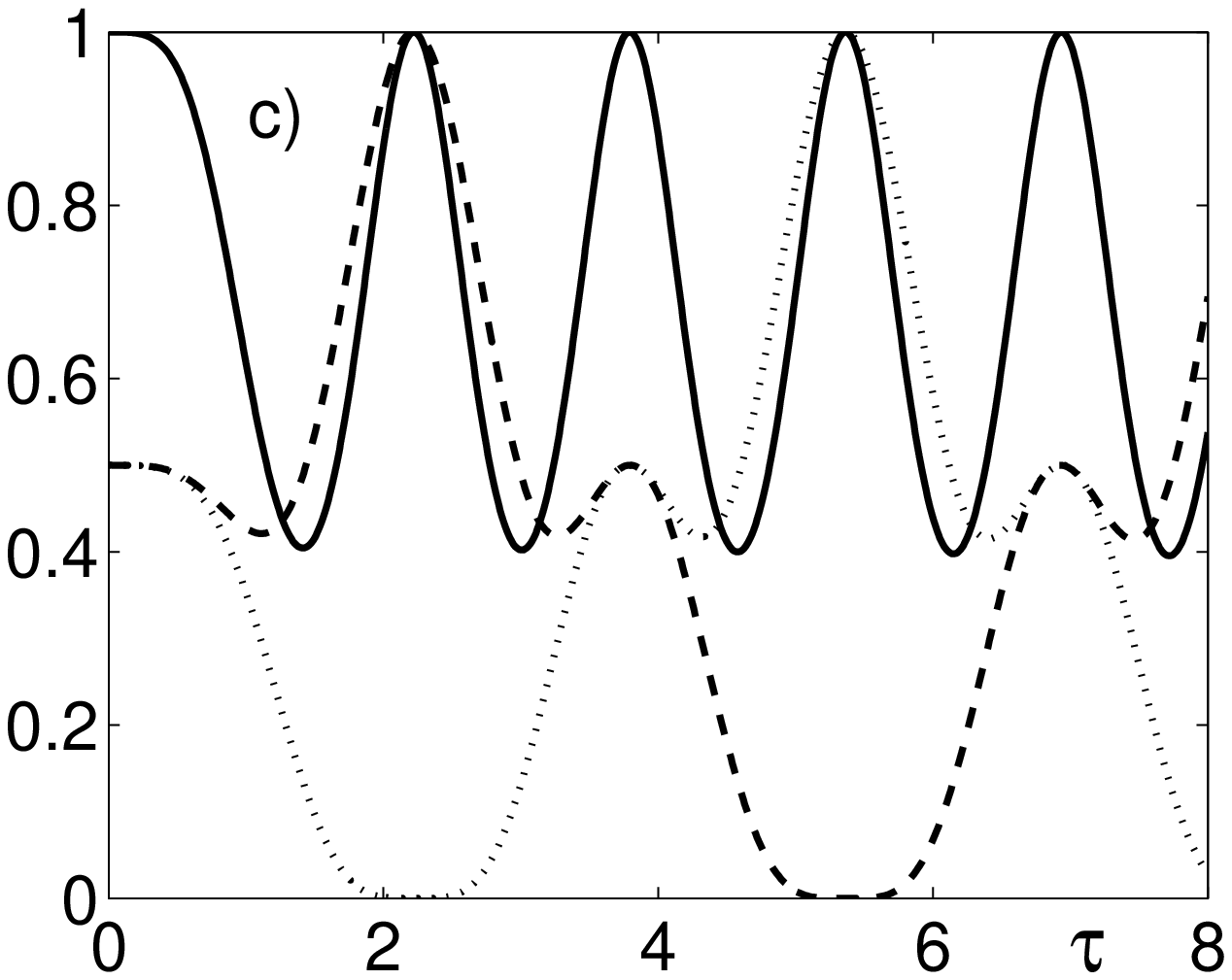}
\includegraphics[width=0.23\textwidth]{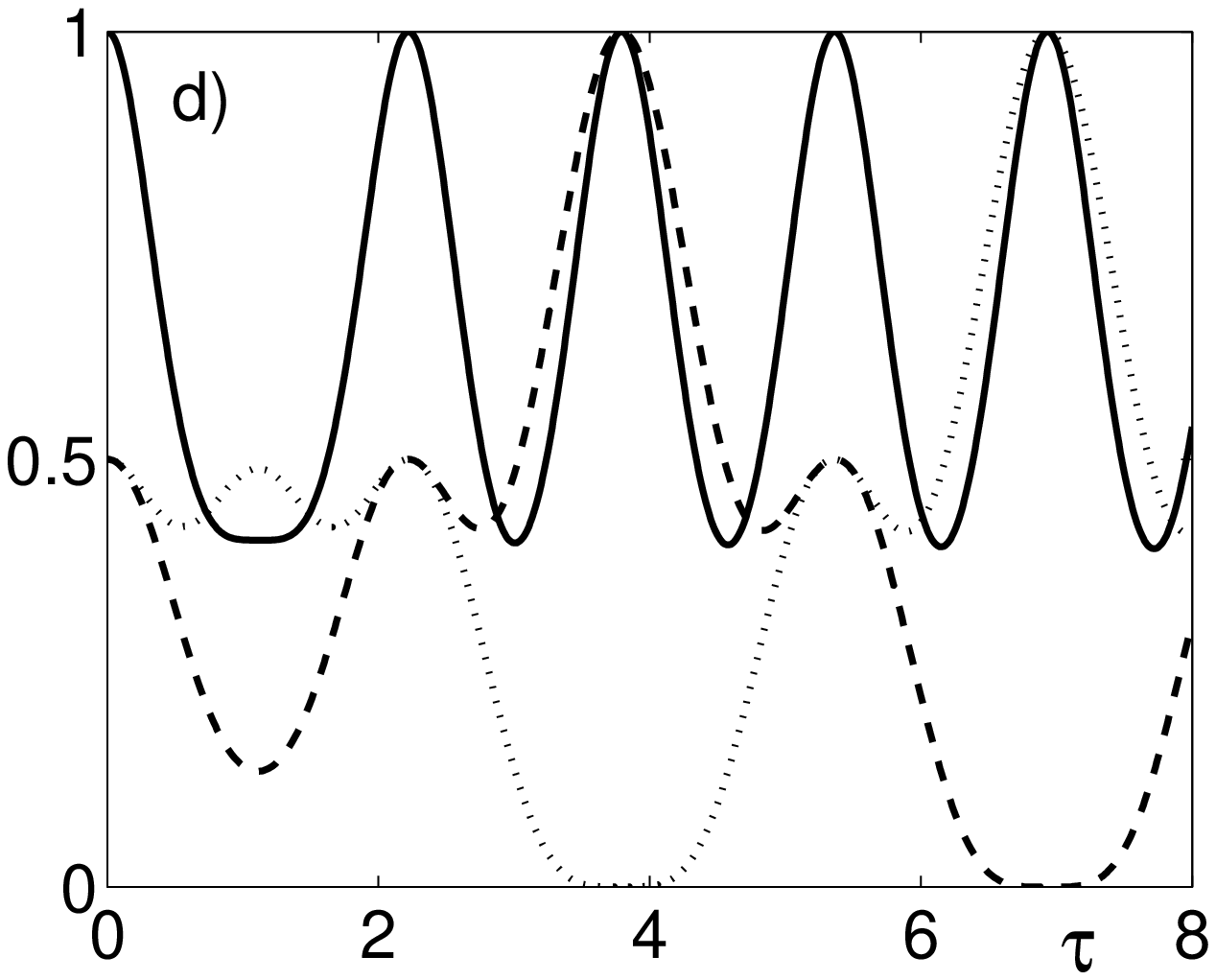}
\caption{\label{fig:fig1}Dynamics for the external field in a GHZ
state: (a) $N^{(c)}$(dashed), $N^{(f)}$(dotted) and $p_e$(solid);
(b) $E^{(\alpha)}$ for atoms (solid), cavity modes (dashed) and
external field (dotted); (c) $\mu^{(a)}$ (solid) and
$F_{\phi}^{(a)}$ with $\phi=0$ (dashed), $\phi=\pi$ (dotted); (d)
$\mu^{(c)}$ (solid) and $F_{\phi}^{(c)}$  with $\phi=-\pi/2$
(dashed), $\phi=+\pi/2$ (dotted).}
\end{figure}

\indent We first illustrate the Hamiltonian dynamics
($\{\tilde{\kappa}_f,\tilde{\kappa}_c,\tilde{\gamma}_a\}\ll 1$)
for the external field prepared in a qubit-like entangled pure
state $\ket{\Psi(0)}_f$, atoms prepared in the lower state
$\ket{ggg}_a$, and cavities in the vacuum state $\ket{000}_c$ . By
choosing $\nu_{JK}(\tau)=0$ if $J\neq K$ and $\nu_{J,J}=g_A$ we
describe single-mode fibers each one propagating a mode of the
entangled radiation.  We show that under optimal conditions for
mode matching it is possible to map $\ket{\Psi(0)}_f$ onto atomic
and cavity mode states for suitable interaction times. Overall we
are dealing with an interacting 9-qubit system, though the input
field will be switched off at a time $\tau_{off}$ such that the
atomic probability of excited state $p_e(\tau)$ reaches the
maximum. Injected field
switch-off can be obtained, e.g., by rotating fiber polarization.\\
\indent In Fig.~\ref{fig:fig1} we show numerical results for the
external field prepared in the GHZ state
$\ket{\Psi(0)}_f=(\ket{000}_f+\ket{111}_f)/\sqrt{2}$. In the time
interval $0<\tau \leq\tau_{off}$ (transient regime) each flying
qubit transfers its excitation to the cavity which in turn passes it
onto the atom (see Fig.~\ref{fig:fig1}a). The cavity mode,
simultaneously coupled to the external field and to the atom,
exchanges energy according to a Tavis-Cummings dynamics at an
effective frequency $g\sqrt{2}$ \cite{Haroche_Raimond,TC}. During
the transient up to time $\tau_{off}=\pi/\sqrt{2}$ the mean photon
number $N^{(c)}(\tau)\equiv\langle \hat{c}^\dag
\hat{c}\rangle(\tau)$ in each cavity completes a cycle. In the same
period the atomic excitation probability $p_e(\tau)$ reaches its
maximum value, while the input field has completely fed the cavity,
i.e., its mean photon number $N^{(f)}(\tau)\equiv\langle
\hat{f}^\dag \hat{f}\rangle(\tau)$ vanishes.
In Fig.~\ref{fig:fig1}b we show that in the transient the atomic
tripartite negativity is always positive and
$E^{(a)}(\tau_{off})=1$, that is the value of the injected GHZ
state. Until $\tau_{off}$ the dynamics maps the whole initial
state $\ket{\Psi(0)}_f\otimes\ket{000}_c\otimes\ket{ggg}_a$ onto
the pure state $\ket{000}_f\otimes\ket{000}_c\otimes
\ket{\Psi(0)}_a$, where $\ket{\Psi(0)}_a$ is obtained from
$\ket{\Psi(0)}_f$ by the correspondence
$\ket{0}_f\rightarrow\ket{g}_a$ and
$\ket{1}_f\rightarrow\ket{e}_a$. This is confirmed in
Fig.~\ref{fig:fig1}c by the time evolution of the purity
$\mu^{(a)}(\tau)=\hbox{Tr}_{a}[\hat{\rho}^2_{\alpha}(\tau)]$ and
the fidelity $F^{(a)}(\tau)={}_{a}
\meanvalue{\Psi(0)}{\hat{\rho}_{a}(\tau)}{\Psi(0)}_{a}$, where
$\hat{\rho}_{a}(\tau)$ is the atomic reduced density operator. As
for the cavity mode dynamics we note that (see
Fig.~\ref{fig:fig1}b,d) the local maximum of
$E^{(c)}(\tau_{off}/2)$ does not correspond to a pure state, i.e.
the initial state $\ket{\Psi(0)}_f$ cannot be mapped onto the
cavity modes during the transient regime. The entanglement is only
partially transferred to the cavity modes nevertheless allowing
the building up of full atomic entanglement later on. This
dynamics is quite different than in \cite{Casagrande} where the
entangled field was mapped onto the cavity modes before the interaction
with the atoms.\\
\indent At the end of the transient regime the external radiation is
turned off and the subsequent dynamics is described by a triple JC
ruled by oscillations at the vacuum Rabi frequency $2g$, hence with
a dimensionless period $\pi$ as shown by cavity mean photon number
and atomic probability in Fig.~\ref{fig:fig1}a. The purities
$\mu^{(a,c)}(\tau)$ in Figs.~\ref{fig:fig1}c,d oscillate at a double
frequency between pure entangled (maximum negativity) and separable
(zero negativity) states. In particular, at times
$\tau_m=\tau_{off}+m\pi$ ($m=0,1,2...)$ the atoms are in the
entangled states $\hat{U}^{(a)}_{\phi}\ket{\Psi(0)}_a$, where
$\hat{U}^{(a)}_{\phi}=\bigotimes_{J}e^{-i\phi\hat{\sigma}_J^{\dag}\hat{\sigma}_{J}}$
is a local phase operator where $\phi=0$ ($\phi=\pi$) applies for
even (odd) values of $m$, that are the peaks of $E^{(a)}(\tau)$ in
Fig.~\ref{fig:fig1}b. At times
$\tau_n=\tau_{off}+(n+\frac{1}{2})\pi$ ($n=0,1,2...$) the cavity
mode states are obtained by applying
$\hat{U}^{(c)}_{\phi}=\bigotimes_{J}e^{-i\phi\hat{c}_J^{\dag}\hat{c}_{J}}$,
where $\phi=-\frac{\pi}{2}$ $(+\frac{\pi}{2})$ for even (odd) values
of $n$, to the state $\ket{\Psi(0)}_c$ derived from
$\ket{\Psi(0)}_f$ by the correspondence
$\ket{0}_f\leftrightarrow\ket{0}_c$ and
$\ket{1}_f\leftrightarrow\ket{1}_c$. By choosing to turn off the
external field at times shorter than $\tau_{off}$ we find a
progressive reduction in the entanglement transfer to the atomic and
cavity subsystems, simulating the effect of non perfect cavity
mirror transmittance. Up to 10\% changes in the value of
$\tau_{off}$, the fidelity $F^{(a)}(\tau_{off})$ remains above
99.9\%. We remark that the state mapping process can be obtained for
any $\ket{\Psi(0)}_f$ written in a generalized Schmidt decomposition
\cite{Dur}, as well as for mixed states as described below.\\
\subsection{Tripartite entanglement sudden death for external field in a Werner state}
\begin{figure}[h]
\begin{tabular}{cc}
\includegraphics[scale=0.30]{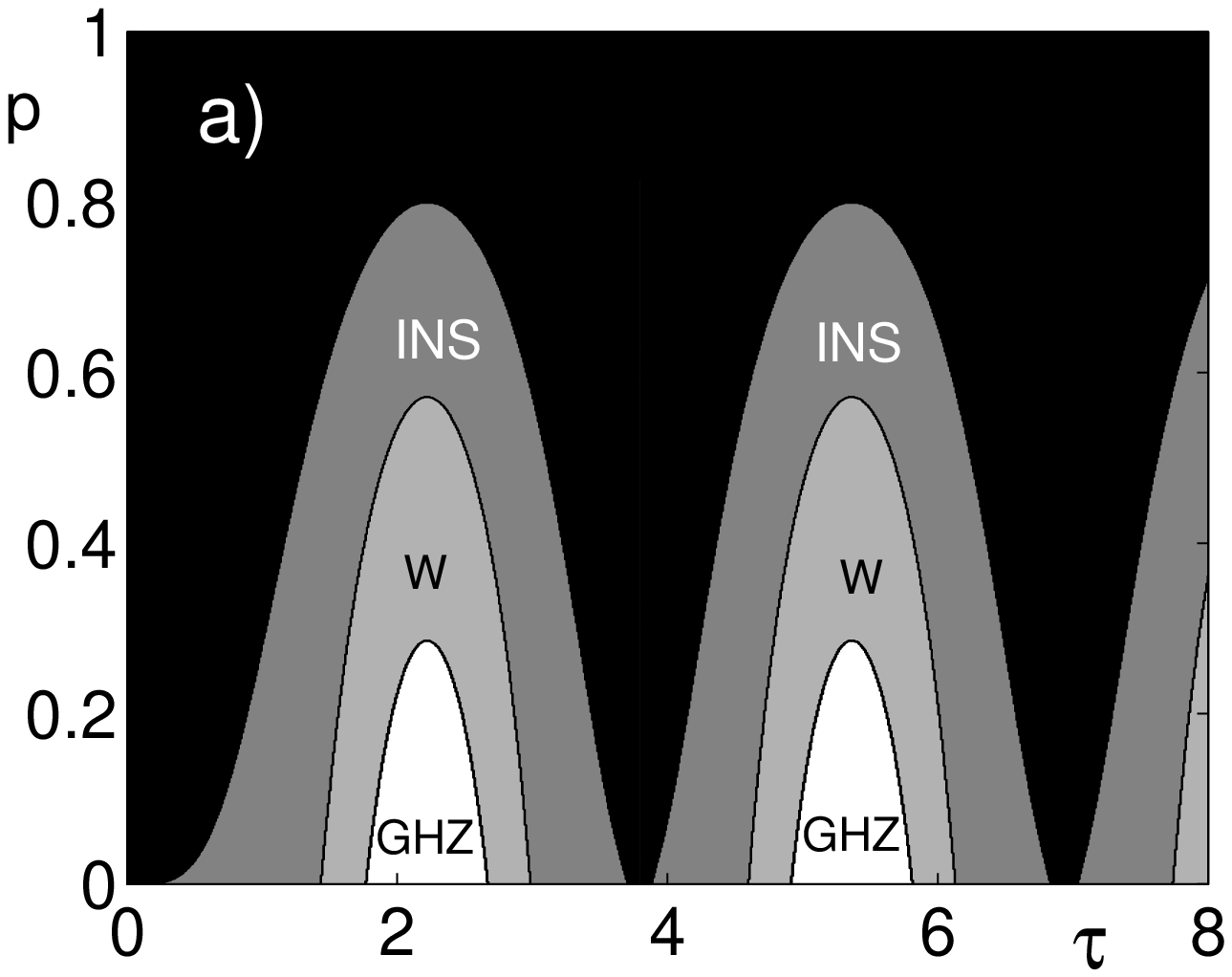}
\includegraphics[scale=0.30]{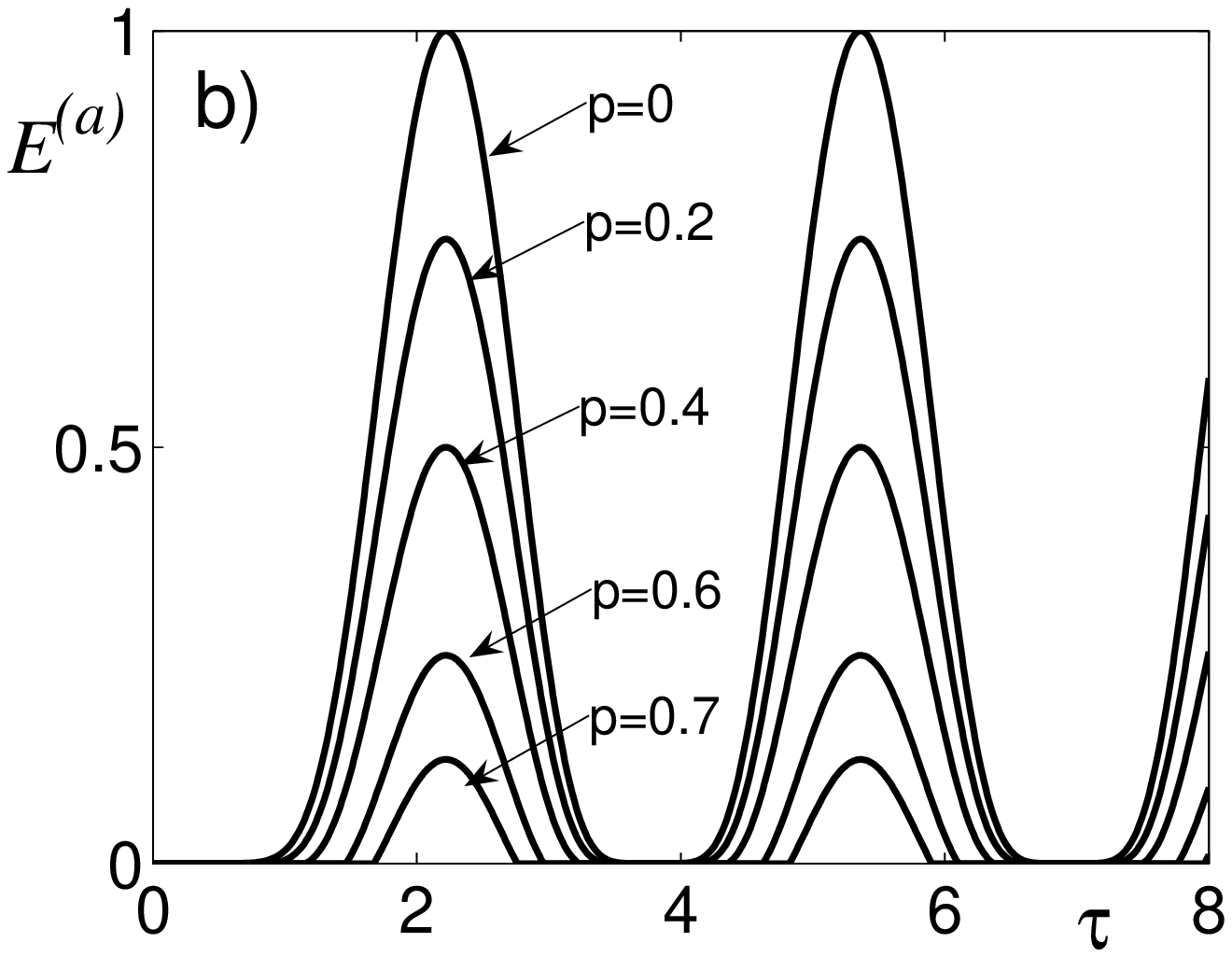}\\
\includegraphics[scale=0.30]{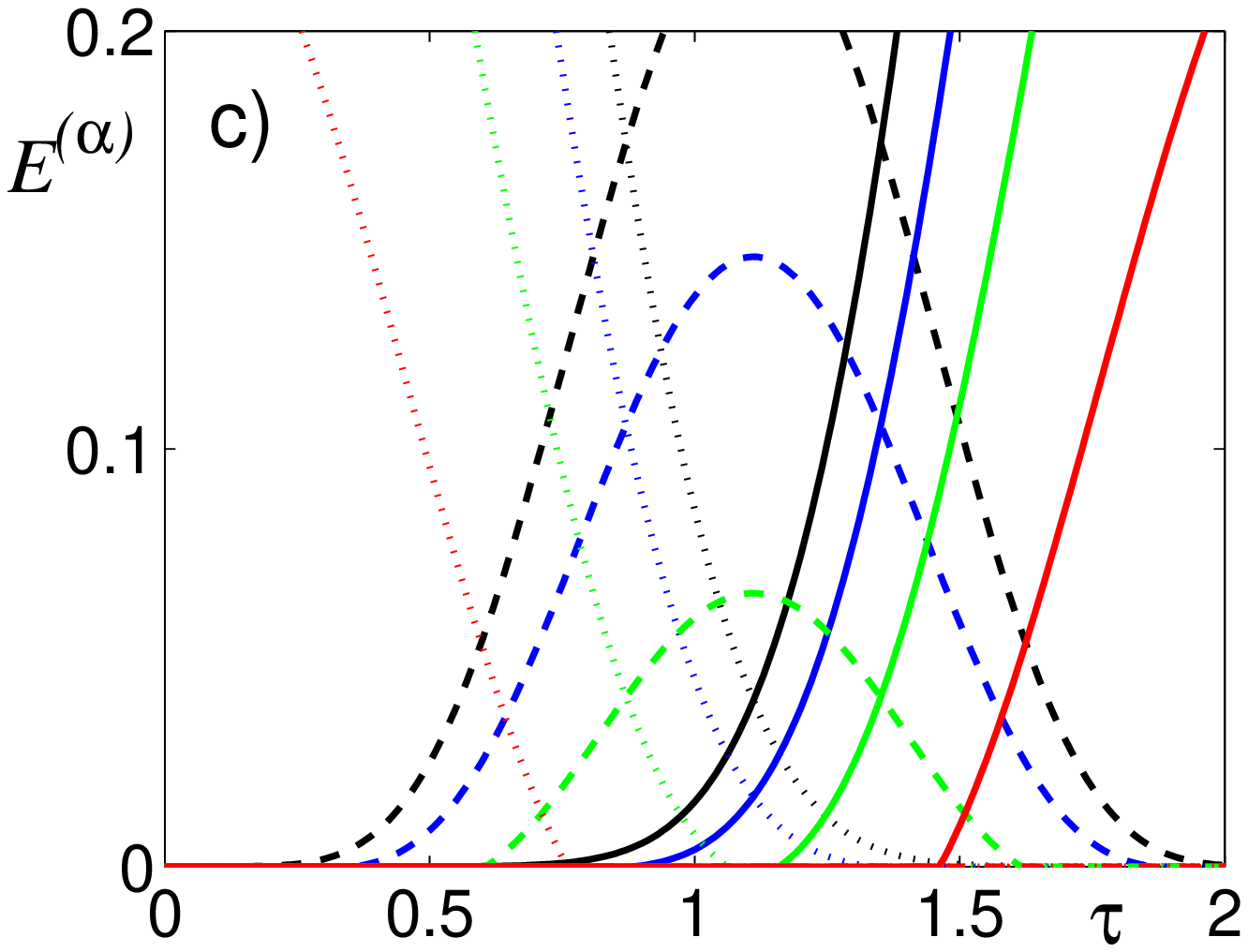}
\includegraphics[scale=0.30]{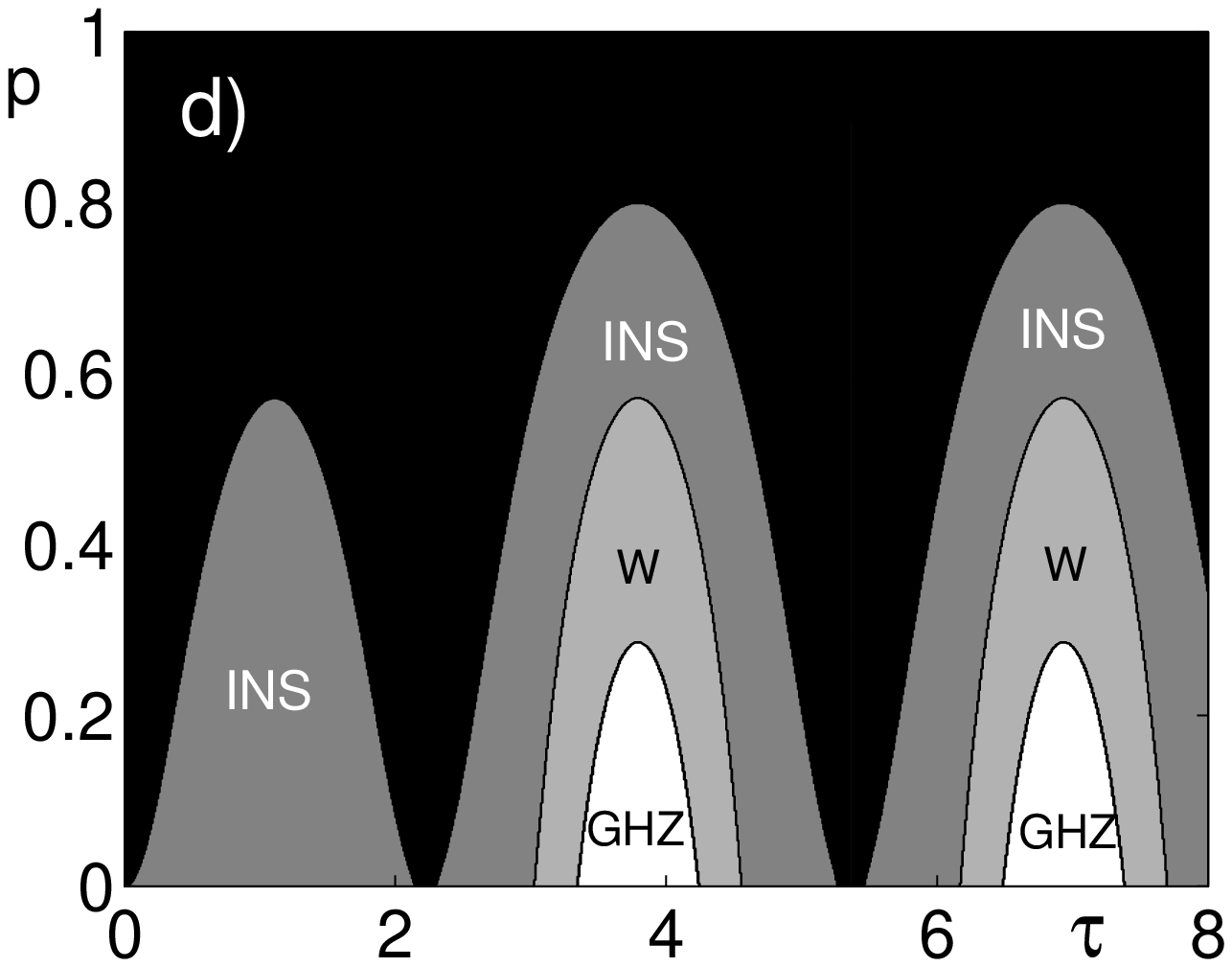}
\end{tabular}
\caption{\label{fig:fig2} ESD/ESB for external field in a GHZ state
mixed with white noise. a) Regions in the plain ($\tau,p$) for
atomic entanglement of type GHZ, W, INS, and fully separable
(black). b) Sections $E^{(a)}(\tau)$ for selected values of $p$. c)
Zoom on $E^{(\alpha)}(\tau_{off}/2)$ for field (dotted), cavity
modes (dashed), and atoms (solid) with $p=0$ (black), $p=0.2$
(blue), $p=0.4$ (green), $p=0.6$ (red). d) Classification for cavity
mode entanglement.}
\end{figure}

\indent For the injected field we consider the Werner state
$\hat{\rho}_f(0)=(1-p)\ketbra{GHZ}{GHZ}+\frac{p}{8} \hat{I}$,
$(0\leq p\leq 1)$, because it is relevant in QI and it is possible
to fully classify its entanglement as a function of parameter $p$.
In fact, for $0\leq p<\frac{2}{7}$ the state belongs to the GHZ
class and to the W class up to $p=\frac{4}{7}$. The tripartite
negativity is positive up to $4/5$, and due to the structure of
$\hat{\rho}_f(0)$, the state is clearly inseparable under all
bipartitions (INS), i.e. it cannot be written as convex
combination of biseparable states. For $4/5\leq p\leq1$ it is
known that the state
is fully separable \cite{Pittenger,Ghune2}.\\
\indent The system dynamics can be divided into a transient and an
oscillatory regime, and the state mapping of $\hat{\rho}_f(0)$ onto
atoms (cavity modes) still occurs at times $\tau_m$ ($\tau_n$). Out
of these times the density matrices of all subsystems lose the form
of a GHZ state mixed with white noise but still preserve invariance
under all permutations of the three qubits and present only one non
vanishing coherence as in $\hat{\rho}_f(0)$. This greatly helps us
in the entanglement classifications in the plane $(\tau,p)$ shown in
Fig.~\ref{fig:fig2}.  In fact, in the regions where
$E^{(\alpha)}(\tau)>0$ but out of W class we can exclude the
biseparability. The fully separability criteria in \cite{Ghune2} are
violated only where $E^{(a)}(\tau)>0$ so that if $E^{(a)}(\tau)=0$
the state may be fully separable or biseparable. Nevertheless, in
the latter case the state should be symmetric and biseparable under
all bipartitions and hence it is fully separable \cite{Kraus}. For
any fixed value of $p$ in the range $0<p<4/5$ we thus show the
occurrence of entanglement sudden death and birth at the boundaries
between fully separable and INS states. In particular, for $0<p<4/7$
we find genuine tripartite ESD and ESB phenomena. Note that, for a
fixed value of $p$, the passage of atomic state during time
evolution from W-class to GHZ-class (or viceversa) entangled states
is permitted by the non-unitarity of the partial trace over
non-atomic degrees of freedom (so that the overall operation on the
initial three qubits is not SLOCC). We also notice that for times
$\tau\geq\tau_{off}$ we can solve exactly the triple JC dynamics,
confirming our numerical results and generalizing \cite{ESDChina} to
mixed states.\\
\indent In Fig.~\ref{fig:fig2}b we see that, for increasing values
of $p$, there is an increase of both the slope of $E^{(a)}(\tau)$
and the time interval of fully separability.
In Fig.~\ref{fig:fig2}c we show in detail the transient dynamics of
the tripartite negativities $E^{(\alpha)}(\tau,p)$ ($\alpha=a,c,f$)
in the crucial region around $\tau_{off}/2$. We consider some values
of $p$ where the atoms exhibit in times different classes of
entanglement. We see that for $p=0.2$, where the input state has GHZ
class entanglement, the ESB of subsystems (c),(a) anticipates the
ESD of (f),(c), and there is an interval around $\tau_{off}/2$ where
all three subsystems are entangled (of INS-type). As $p$ grows,
hence the initial state becomes more noisy, the effects of ESD occur
earlier and those of ESB later. For $p=0.4$, involving W-class
entanglement, only at most two subsystems are simultaneously
entangled (first (f),(c) and then (c),(a)). For $p=0.6$, involving
only entanglement of INS-type, the cavity modes do not entangle at
all (see Fig.~\ref{fig:fig2}d). They physically mediate the
discontinuous entanglement transfer from (f) to (a), where for
$p\rightarrow 4/5$ the time interval without any
entanglement increases while the entanglement level vanishes.\\
\subsection{Effect of dissipation on state mapping} \indent In the
perspective of experimental implementation for QI purposes an
important issue is the effect of dissipation on both state mapping
and entanglement transfer. For external field prepared in a GHZ pure
state we first evaluated the effect of cavity decay rates in the
range $0<\tilde{\kappa}_c\leq0.5$ for negligible values of all other
decay rates. We consider as function of $\tilde{\kappa}_c$ the
behavior of the fidelities $F^{(\alpha)}(\tau_{m,n})$ and the
tripartite negativities $E^{(\alpha)}(\tau_{m,n})$ ($\alpha=c,f$) at
the first peaks $(m=n=0)$.
\begin{figure}[h]
\begin{tabular}{cc}
\includegraphics[scale=0.40]{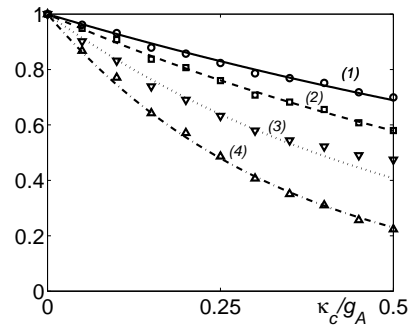}
\end{tabular}
\caption{\label{fig:fig3} Effect of cavity mode dissipation. At
the first peaks $\tau_{m,n}$ with $m=n=0$ we consider the
tripartite negativities as function of
$\tilde{\kappa}_c\equiv\frac{\kappa_c}{g_A}$: $F^{(a)}(\tau_0)$
(1), $E^{(a)}(\tau_0)$ (2), $F^{(c)}(\tau_0)$ (3),
$E^{(c)}(\tau_0)$ (4).}
\end{figure}
In Fig.~\ref{fig:fig3} we see that the above functions of
$\tilde{\kappa}_c$ can be well fitted by exponential functions,
whose decay rates for the atomic subsystem are
$\beta^{(a)}_{F}=0.75$, $\beta^{(a)}_{E}=1.09$, and for the cavity
modes $\beta^{(c)}_{F}=1.80$, $\beta^{(c)}_{E}=2.94$. As expected,
quantum state mapping and entanglement transfer are by far more
efficient onto atomic than cavity qubits. For instance, if
$\tilde{\kappa}_c = 0.1$ we obtain a state mapping onto the atoms
(cavity modes) with a
fidelity of $\cong0.93$ ($\cong0.83$).\\
\indent We can now add the further dissipative effect of atomic
decay. For instance we find that, for an atomic decay rate
$\tilde{\gamma}_a=0.03$ and in the presence of cavity decay with a
rate $\tilde{\kappa}_c = 0.1$, the fidelity of the atomic (cavity
mode) subsystem reduces by $4.4\%$ ($8.9\%$).\\
\indent Finally, we evaluate the effect of losses in the fibers used
to inject the external field into each cavity. Clearly, this effect
is relevant only up to the time $\tau_{off}=2.22$. We evaluated the
effects of fiber decay rates $\tilde{\kappa}_f$ up to 1.0 for
negligible values of atomic and cavity decay rates
($\tilde{\kappa}_c<<1$, $\tilde{\gamma}_a<<1$) (see
Fig.~\ref{fig:fig4}). We show the effect of parameter
$\tilde{\kappa}_f$ on cavity field mean photon number
$N^{(c)}(\tau_{off}/2)$ and atomic excitation probability
$p_e(\tau_{off})$ and we see that the amount of energy transferred
to the atoms and to the cavity modes decreases exponentially for
increasing values of $\tilde{\kappa}_f$; the decay rates are $\cong
0.42$ and $\cong 0.82$, respectively. Also the behavior of the
tripartite negativity $E^{(a)}(\tau_{0})$ and fidelity
$F^{(a)}(\tau_{0})$ at the first peak  can be described versus
$\tilde{\kappa}_f$ by exponential functions whose decay rates are
$\cong1.51$ and $\cong 1.95$, respectively.
\begin{figure}[h]
\begin{tabular}{cc}
\includegraphics[scale=0.40]{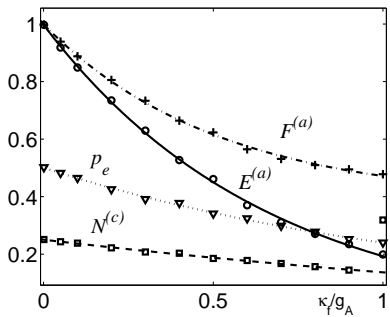}
\end{tabular}
\caption{\label{fig:fig4} Effect of fiber mode decay rate
$\tilde{\kappa}_f\equiv\kappa_f/g_A$ for $\tilde{\kappa}_c<<1$,
$\tilde{\gamma}_a<<1$. We evaluate at time $\tau_{off}$ the atomic
tripartite negativity $E^{a}$(solid), the fidelity
$F^{a}$(dash-dot), and the atomic probability $p_e$(dot), and at
time $\tau_{off}/2$ the cavity mean photon number
$N^{(c)}$(dash).}
\end{figure}
\section{State mapping and tripartite entanglement transfer for multi-mode coupling}
\begin{figure}[h]
\begin{tabular}{cc}
\includegraphics[scale=0.30]{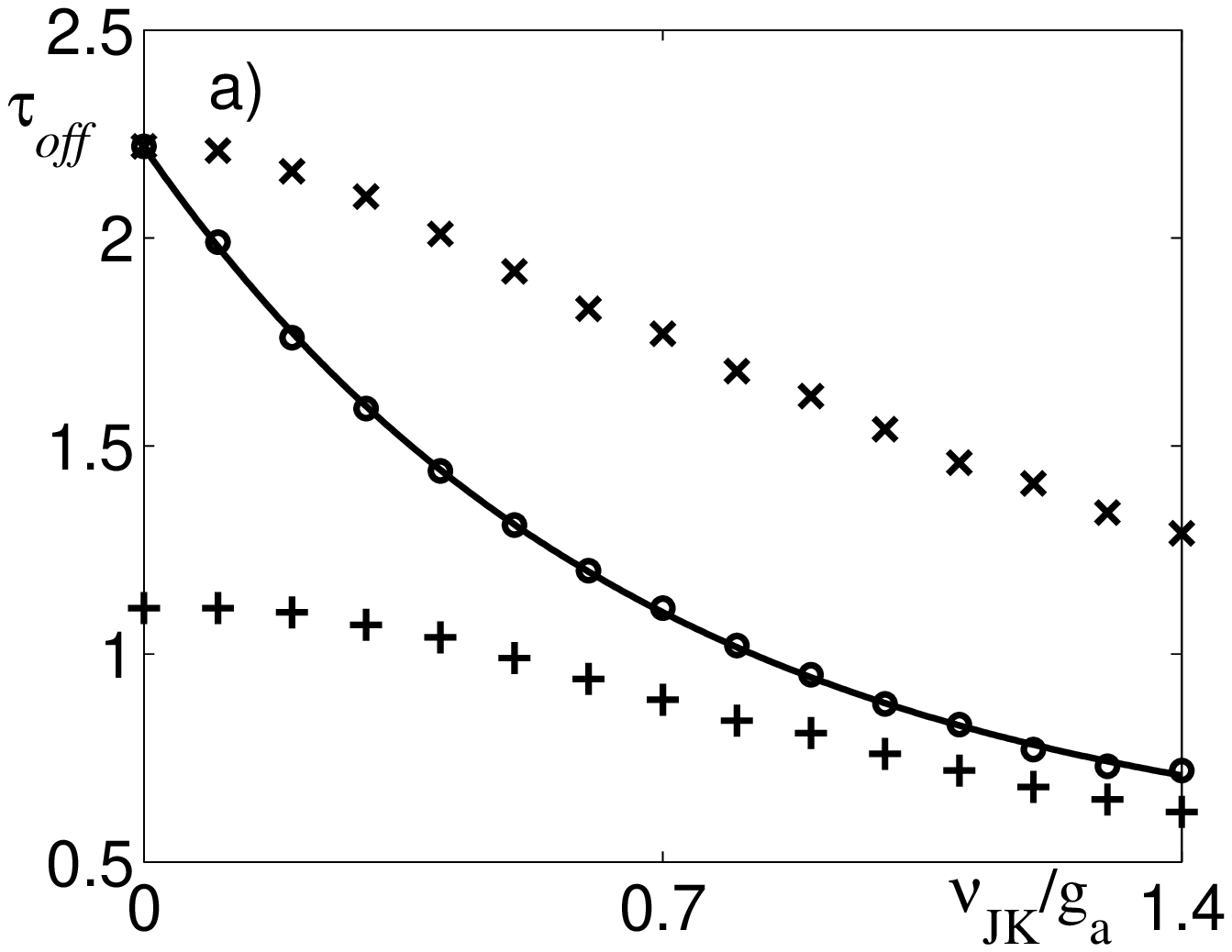}\\
\includegraphics[scale=0.27]{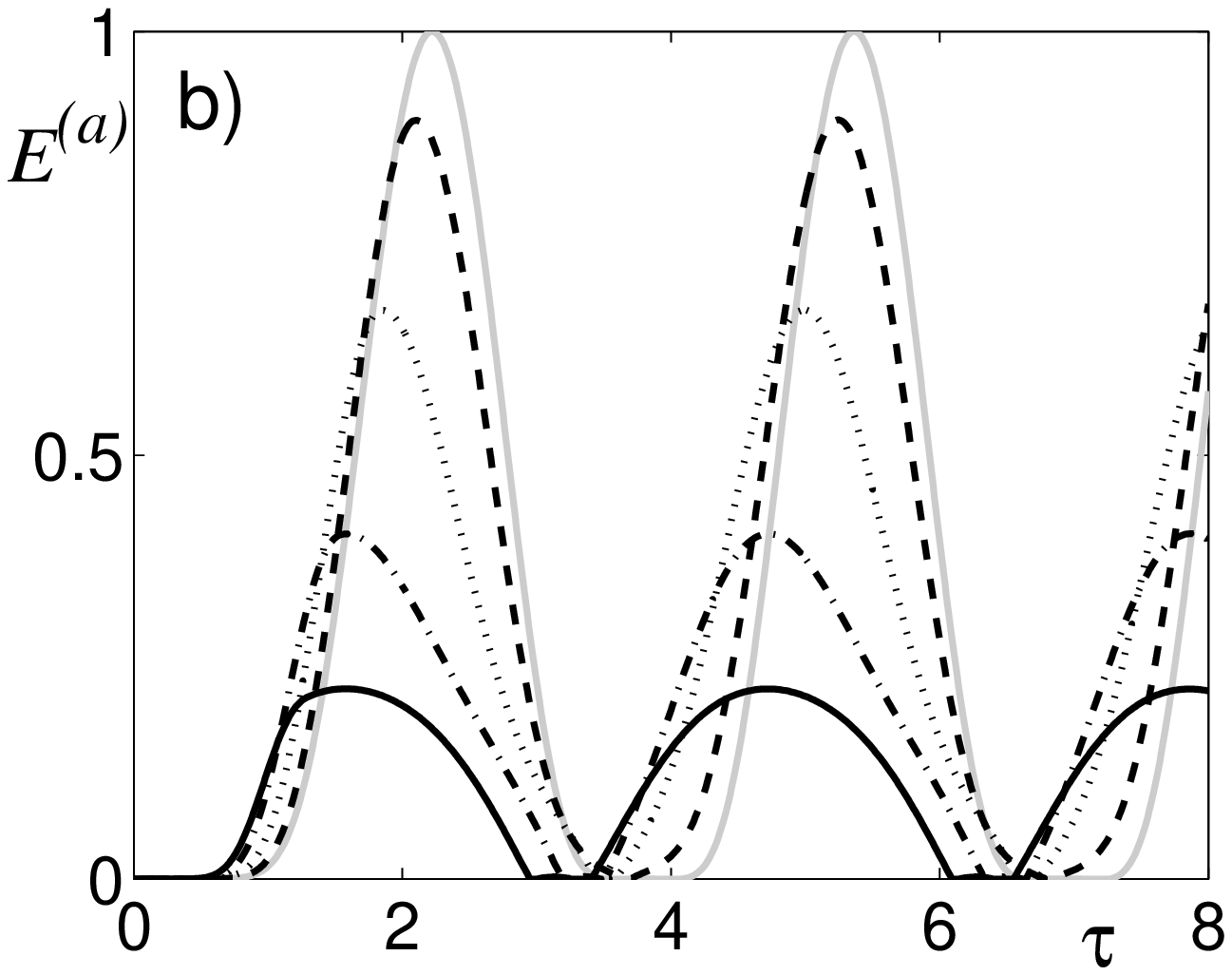}
\includegraphics[scale=0.27]{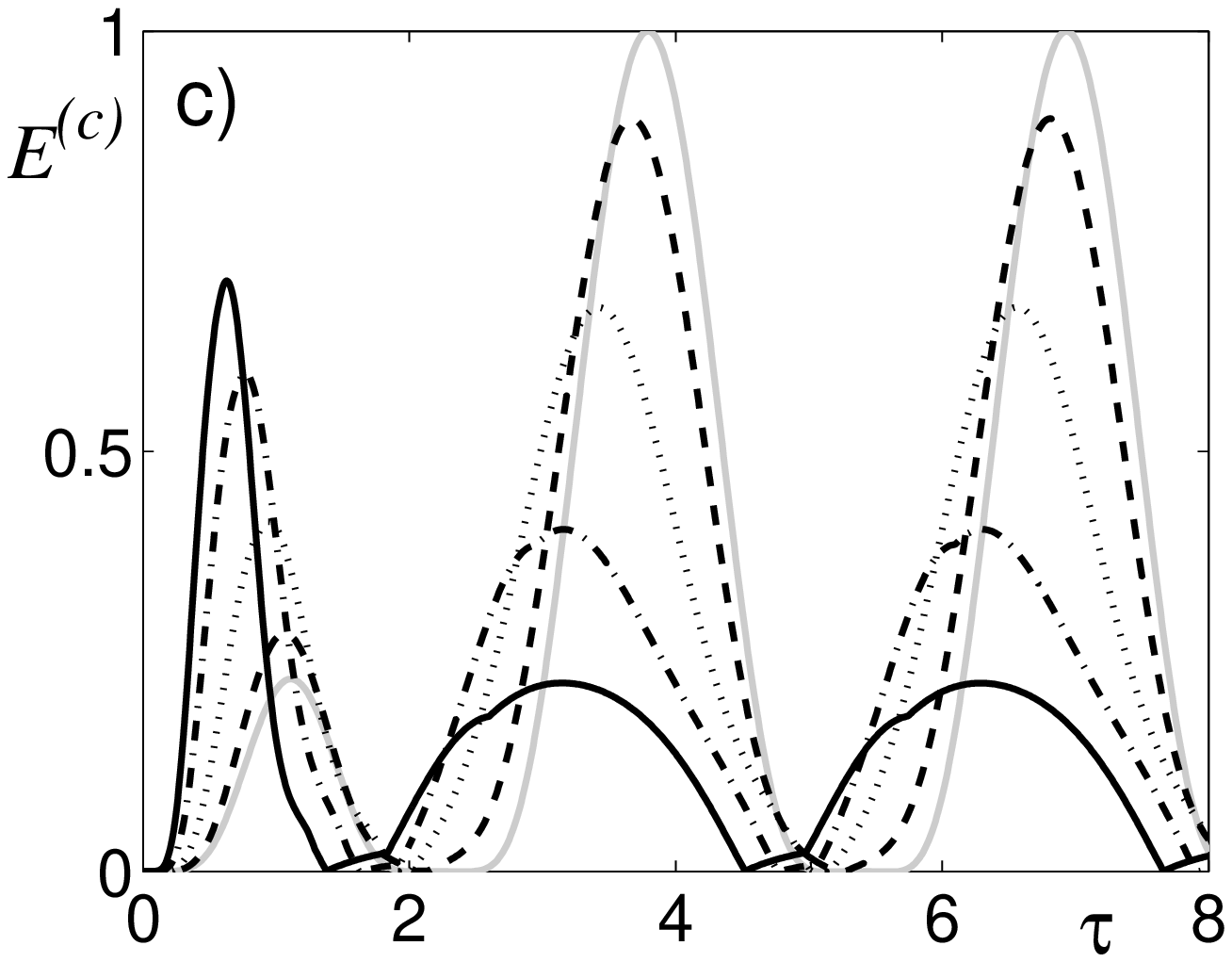}\\
\includegraphics[scale=0.27]{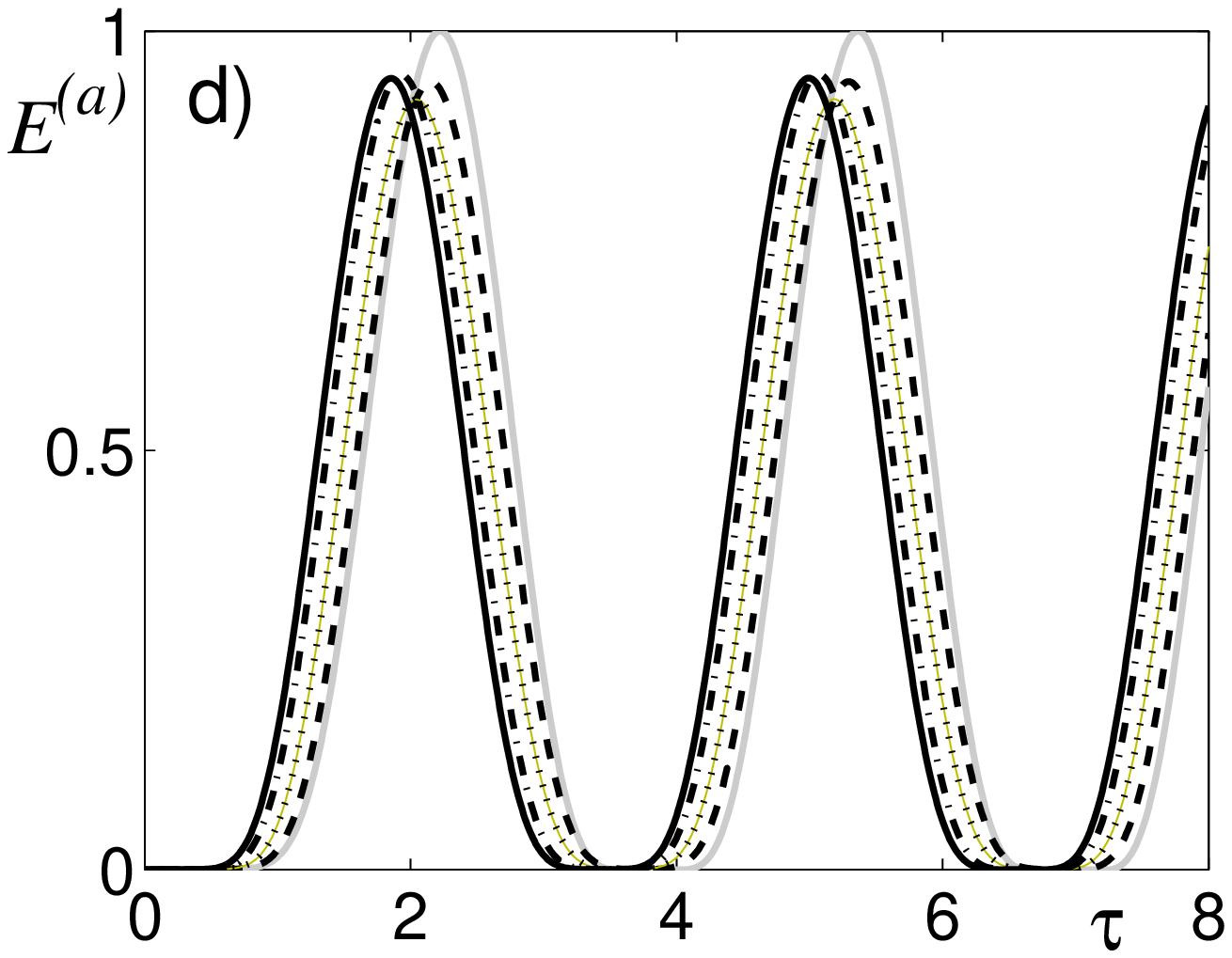}
\includegraphics[scale=0.27]{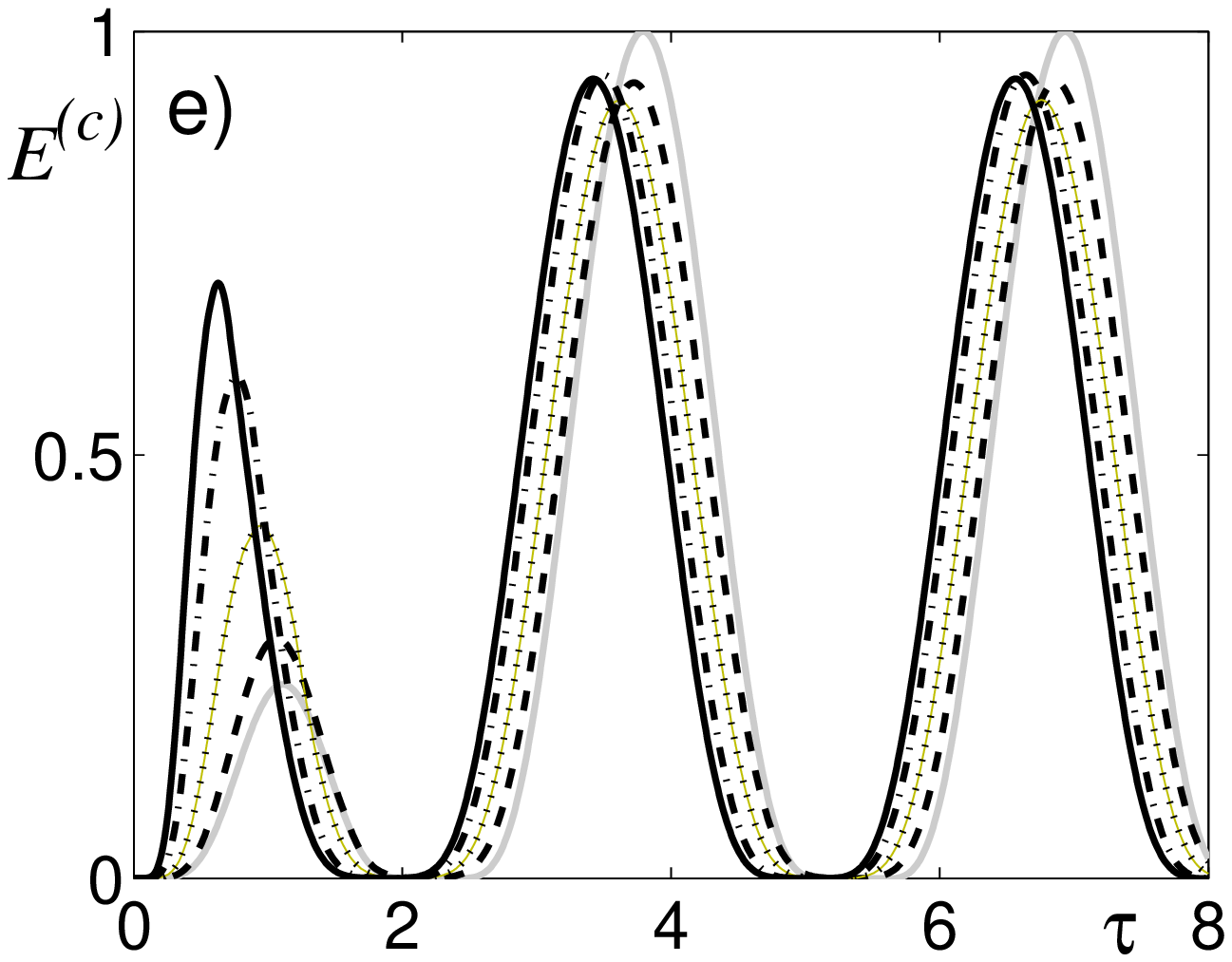}\\
\includegraphics[scale=0.27]{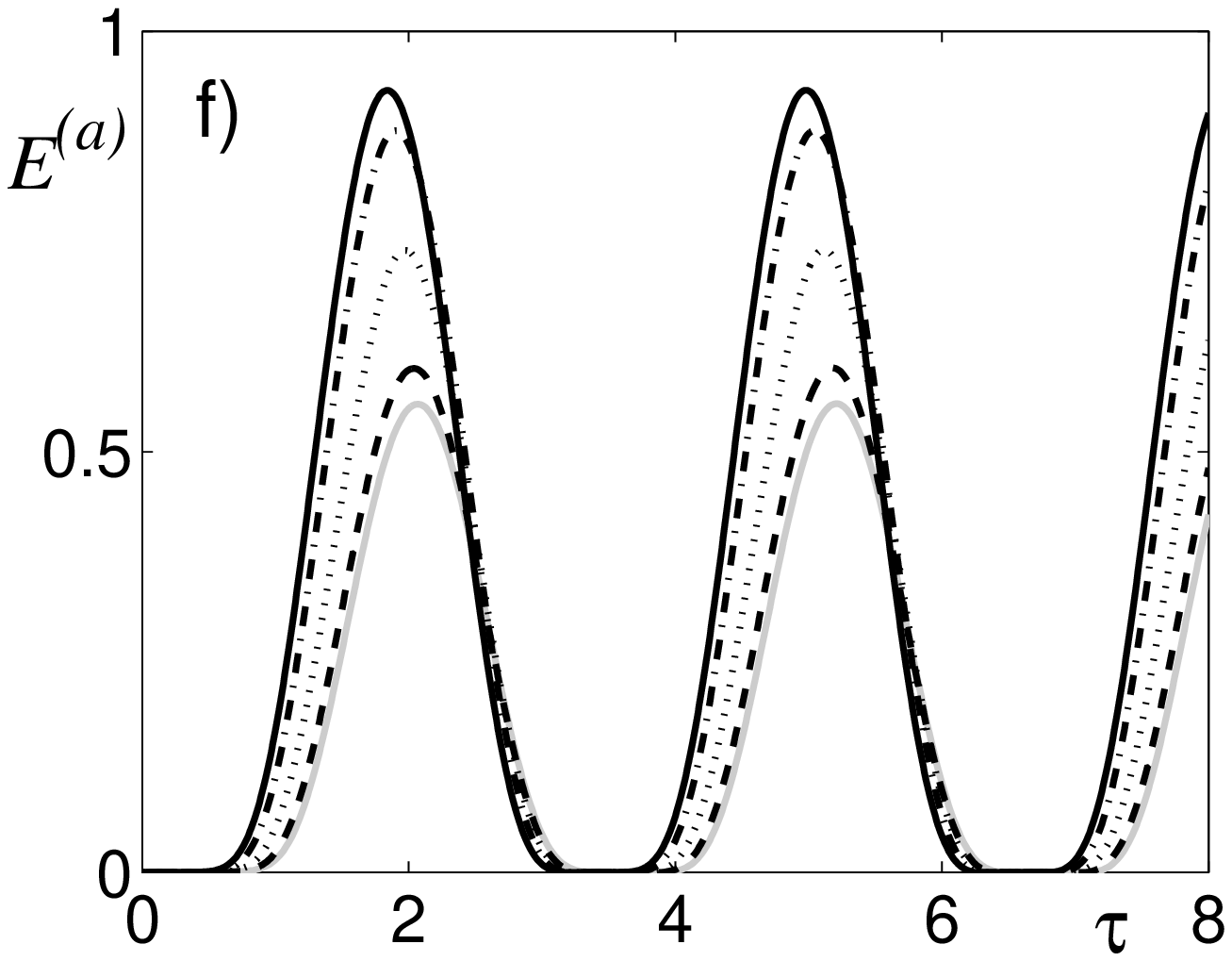}
\includegraphics[scale=0.27]{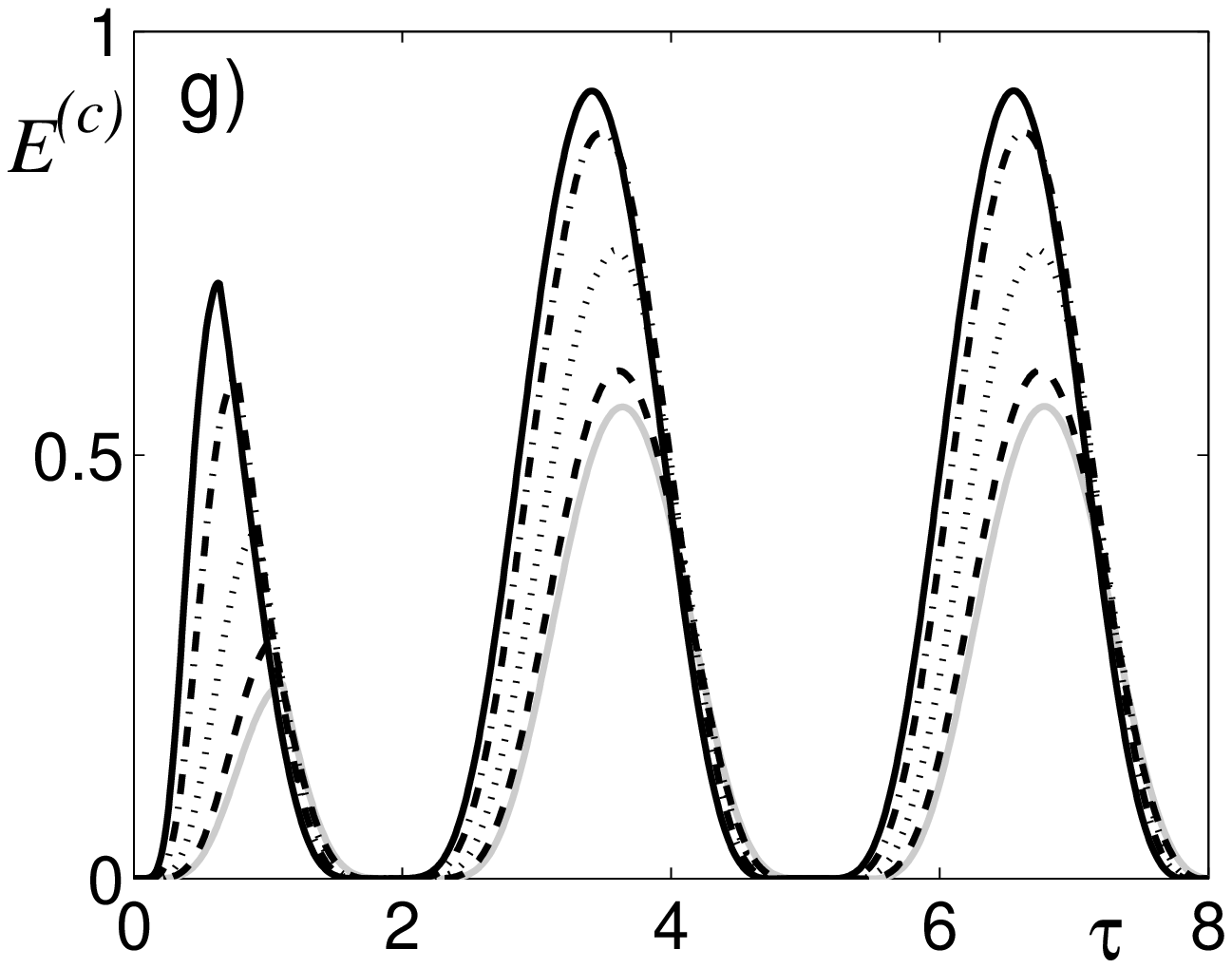}
\end{tabular}
\caption{\label{fig:fig5} Effect of multimode coupling. a)
Dependence of $\tau_{off}$ on the coupling constants
$\tilde{\nu}_{J,K}$ for different choices of switching-off the
external field: maximum of $p_{e}(\tau)$ (o), minimum of
$N^{(f)}(\tau)$ (x), and maximum of $N^{(c)}(\tau)$ (+). Tripartite
negativities $E^{(\alpha)}$ ($\alpha=a,c$) for $\tilde{\nu}_{J,K}=0$
(solid gray), 0.3 (dashed), 0.6 (dotted), 1.0 (dashed-dotted), and
1.4 (solid black): b,c)  $\tau_{off}$ in the maximum of
$p_{e}(\tau)$; d,e) $\tau_{off}$ in the minimum of $N^{(f)}(\tau)$;
f,g) $\tau_{off}$ in the maximum of $N^{(c)}(\tau)$.}
\end{figure}

\indent Finally, we consider multi-mode coupling of the external
field to each cavity mode. For simplicity we choose equal coupling
constants $\tilde{\nu}_{J,K}\equiv\nu_{J,K}/g_A\neq 0$ if $K\neq J$
and we consider values in the range $0-1.4$. In the transient regime
the dynamics is sharply modified with respect to the case of single
mode fiber shown in Fig.~\ref{fig:fig1}. By increasing the values of
$\tilde{\nu}_{J,K}$ the period of energy exchange decreases from
$2\pi/\sqrt{2}$ to $\cong2.6$. The maximum of cavity mode mean
photon number grows up to $N^{(c)}\cong 0.41$ whereas the maximum of
atomic excitation probability decreases to $p_{e}\cong 0.24$. The
external field mean photon number does not vanish but it reaches a
minimum, that can be always found between the two maxima of
$N^{(c)}(\tau)$ and $p_{e}(\tau)$, such that $ 0.002<N^{(f)}<0.02$
changing $\tilde{\nu}_{J,K}$ from 0.1 to 1.4. We investigate the
differences in the entanglement transfer for three selections of
switching-off time $\tau_{off}$ corresponding to the maximum of
$p_{e}(\tau)$, the minimum of $N^{(f)}(\tau)$, and the maximum of
$N^{(c)}(\tau)$. In Fig.~\ref{fig:fig5}a we show the dependence of
$\tau_{off}$ on $\tilde{\nu}_{J,K\neq J}$. Switching off the
external field at times $\tau_{off}$ corresponding to the maxima of
$p_{e}(\tau)$, as in the previous case with single-mode fibers, we
find (Fig.~\ref{fig:fig5}b,c) that the maxima of tripartite
negativities $E^{(\alpha)}(\tau)$ after the transient regime reduce
for increasing values of $\tilde{\nu}_{J,K}$ for both atomic and
cavity mode subsystems.
If we consider $\tau_{off}$ corresponding to the minimum of
$N^{(f)}(\tau)$ (Fig.~\ref{fig:fig5}d,e) we observe a small
reduction of the peak values of $E^{(\alpha)}(\tau)$. Finally, if we
turn off the external field at the first maximum of the cavity field
mean photon number we note that by increasing the values of
$\tilde{\nu}_{J,K}$ it is possible to improve the entanglement
transfer (Fig.~\ref{fig:fig5}f,g). The peak value of tripartite
negativity grows up to $\cong0.93$ for $\tilde{\nu}_{J,K}=1.4$ and
the fidelity up to $\cong0.95$ for both subsystems (a) and (c). We
remark that these values cannot be significantly increased for
larger values of $\tilde{\nu}_{J,K}$. In conclusion, for all the
above choices of switching-off time $\tau_{off}$ we observe that, by
increasing the values of $\tilde{\nu}_{J,K}$, the amount of
entanglement that can be transferred to the cavity modes in the
transient regime also increases. This is due to the fact that the
amount of energy transferred to each cavity mode increases: in fact,
the peak value of $N^{(c)}(\tau)$ progressively grows up from $\cong
0.25$ to $\cong 0.41$. Nevertheless, multimode coupling for larger
values of $\tau_{off}$ results in a less favorite condition
for entanglement transfer.\\
\section{conclusions}
 \indent In this paper we have addressed the transfer of quantum
information and entanglement from a tripartite bosonic system to
three localized qubits through their environments, also in the
presence of external environments. We considered an implementation
in the optical regime of CQED based on CV photon number entangled
fields and atomic qubits trapped in one-sided optical cavities,
where the radiation modes can couple to cavity modes by optical
fibers.\\
\indent In the nine-qubit transient regime the quantum state is
mapped from tripartite entangled radiation to tripartite atomic
system via the cavity modes. After the transient we switch off the
external field. The subsequent triple JC dynamics, that we solved
analytically (numerically) for pure (mixed) input states, shows how
the effect of mapping can further affect the atom-cavity six qubits
system. Its relevance is in the possible manipulation for QI
purposes of entanglement stored in separate qubits of atomic or
bosonic nature. Hence the interest to put quantitative limits
dictated by cavity, atomic and fiber mode decays, that we evaluated
at the times where the transfer protocol is optimal.\\
\indent In the case of a GHZ input state mixed with white noise, we
provide a full characterization of the separability properties of
the tripartite subsystems. We can then show the occurrence of
entanglement sudden death effects at the tripartite level, deriving
the conditions for the repeated occurrence of discontinuous exchange
of quantum correlations among the tripartite subsystems. This is an
issue of fundamental interest as well as worth investigating for all
applications in quantum information processing, remarkably computing
and error correction, where disentanglement, which may be faster
than decoherence, has to be carefully controlled.\\
\indent An extension and comparison to other types of entangled
qubit-like input fields and experimentally available CV fields
\cite{CV} will be presented elsewhere \cite{Bina}.
\acknowledgements
This work has been partially supported by the CNR-CNISM convention.

\end{document}